\documentclass[prd,aps,twocolumn,superscriptaddress,preprintnumbers,nofootinbib]{revtex4-1}
\usepackage{pslatex}
\usepackage[pdftex]{graphicx}
\usepackage{graphicx, epsfig}
\usepackage[dvipsnames]{xcolor}
\usepackage{hyperref}
\usepackage{mathrsfs}
\usepackage{amsmath}
\usepackage{enumerate}
\usepackage{bm}

\usepackage{float}
\usepackage{ulem}

\newcommand{\be}{\begin{equation}}
\newcommand{\ee}{\end{equation}}
\newcommand{\bea}{\begin{eqnarray}}
\newcommand{\eea}{\end{eqnarray}}
\newcommand{\ba}{\begin{eqnarray}}
\newcommand{\ea}{\end{eqnarray}}

\newcommand{\gapp}{\mathrel{\raise.3ex\hbox{$>$}\mkern-14mu
              \lower0.6ex\hbox{$\sim$}}}
\newcommand{\lapp}{\mathrel{\raise.3ex\hbox{$<$}\mkern-14mu
              \lower0.6ex\hbox{$\sim$}}}

\begin{document}
\title{Probing new physics for $(g-2)_\mu$ and gravitational waves }

\author{Ruiyu Zhou}
\affiliation{
	Department of Physics, Chongqing University, Chongqing 401331, China
}
\author{Ligong Bian}
\email{lgbycl@cqu.edu.cn}

\affiliation{
	Department of Physics, Chongqing University, Chongqing 401331, China
}

  \author{Jing Shu}
    \email{ jshu@mail.itp.ac.cn}
    \affiliation{CAS Key Laboratory of Theoretical Physics, Insitute of Theoretical Physics, Chinese Academy of Sciences, Beijing 100190, China}
    \affiliation{School of Physical Sciences, University of Chinese Academy of Sciences, Beijing 100049, China}
    \affiliation{CAS Center for Excellence in Particle Physics, Beijing 100049, China}
    \affiliation{School of Fundamental Physics and Mathematical Sciences, Hangzhou Institute for Advanced Study, University of Chinese Academy of Sciences, Hangzhou 310024, China}
    \affiliation{Center for High Energy Physics, Peking University, Beijing 100871, China}
    \affiliation{International Center for Theoretical Physics Asia-Pacific, Beijing/Hanzhou, China}
\date{\today}

\begin{abstract}

We study the possibility of probing new physics accounting for $(g-2)_\mu$ anomaly and gravitational waves with pulsar timing array measurements. The model we consider is either a light gauge boson or neutral scalar interacting with muons. We show that the parameter spaces of dark $U(1)$ model with kinetic mixing explaining $(g-2)_\mu$ anomaly can realize a first-order phase transition, and the yield-produced gravitational wave may address the common red noise observed in the NANOGrav 12.5-yr dataset.

\end{abstract}

\maketitle

\section{Introduction}

The noticeable discrepancy between the experimental observations and the Standard Model (SM) predictions of the muon anomalous magnetic moment ($g-2$) 
lasts for more than a decade. The observed value of $a_{\mu}=(g-2)_\mu/2$ at Brookhaven National Laboratory (BNL) is~\cite{Bennett:2006fi, Tanabashi:2018oca}
\begin{equation}
a_{\mu}^{{\rm exp}} \ = \ (11659209.1\pm 6.3)\times10^{-10}\;.\label{eq:a-exp-val}
\end{equation}
More recently, the Muon $g-2$ experiment at Fermilab published their first result 
\begin{equation}
a_{\mu}^{{\rm exp}} \ = \ (11659204.0\pm 5.4)\times10^{-10}. \label{eq:a-exp-fermi--val}
\end{equation}
However, the SM prediction is~\cite{Tanabashi:2018oca,Blum:2013xva}
\begin{equation}
a_{\mu}^{{\rm SM}} \ = \ (11659183.0\pm 4.8)\times10^{-10} \, , \label{eq:a-SM-val}
\end{equation}
 Therefore, the discrepancy between experiments and theoretical prediction is around $4.2 \sigma$ level, 
\begin{equation}
\Delta a_{\mu}=(25.1\pm5.9)\times10^{-10} \,,
\label{eq:a_mu_val}
\end{equation}
The uncertainty of the QCD calculation of the $(g-2)_\mu$ requires for the continuous effort of the community. The study of Ref.~\cite{Chao:2021tvp} state that the hadronic light-by-light contribution cannot address the anomaly.

This longstanding anomaly motivates a lot of continuous theoretical explanations with variants of new physics models. 
For comprehensive reviews, see Refs.~\cite{Jegerlehner:2009ry,Lindner:2016bgg}. We are particularly interested in two possibilities that have been broadly considered to address this anomaly, with either a light gauge boson (or neutral scalar) interacting with muons. 
In particular, we consider a dark U(1) model and a dark $Z_3$ scalar. 
We note that for the explanation of the $(g-2)_\mu$ anomaly with a light gauge boson of $U(1)_{\mu-\tau}$,  the required gauge coupling is of order $\mathcal{O}(10^{-4}-10^{-3})$ for $\mathcal{O}(1-10^2)$ MeV light gauge boson and is too small to trigger a first-order phase transition (FOPT).
In the meantime, we notice that
the NANOGrav Collaboration reports a strong evidence of a stochastic common-spectrum process~\cite{Arzoumanian:2020vkk} in the 12.5-yr data set. 
Among various gravitational wave background explanations of the process, the phase transition occurring at low temperature of around $T_\star\sim \mathcal {O}(1-10)$ MeV can successfully fit the dataset~\cite{Nakai:2020oit,Addazi:2020zcj,Vaskonen:2020lbd,Ratzinger:2020koh,Bian:2020bps}. Therefore, it is of great interest to study the possible common new physics origin for the $g-2$ anomaly together with pulsar timing arrays experiments.

\section{$(g-2)_\mu$ explanations}

We consider a dark U(1) model, which connect with visible sector through $U(1)_Y\times U(1)_D$ kinetic mixing~\cite{Holdom:1985ag}, with the interaction between the dark gauge boson and electromagnetic current $J_\mu^{\rm em}=\Sigma_f Q_f \bar{f}\gamma_\mu f$ being~\cite{Davoudiasl:2012ag} 
\begin{equation}
\mathcal{L}_{int}=-e\epsilon A_\mu^\prime J_{em}^\mu\;, 
\end{equation}
The dark gauge boson can contribute to the anomalous magnetic moment of lepton radiatively, with~\cite{Pospelov:2008zw,Fayet:2007ua,Leveille:1977rc}
\begin{equation}
\Delta a^{A^\prime}_{\ell} \ = \
\frac{\alpha\epsilon^{2}}{2\pi}\int_{0}^{1} {\rm d}z \frac{2z(1-z)^2 m_\ell^2}{z m_V^2+(1-z)^2 m_\ell^2}\,,
\label{eq:g-2_v}
\end{equation}
with $\alpha=e^2/(4\pi)$ being the fine structure constant,  
$\ell$ being the lepton and $A'$  running in the triangle one-loop $g-2$ diagram. 
The dark gauge boson gets mass after the spontaneous symmetry of the $U(1)_D$ after the phase transition, which can be first-order for moderate scalar couplings and gauge couplings. The bosonic part of the Lagrangian before spontaneous symmetry breaking is given by,
\begin{equation}
\label{Lag1}
{\cal L} =  |D_\mu S|^2 - \frac{1}{4} F^\prime_{\mu\nu}F^{\prime\mu\nu} - V(S)\;,
\end{equation}
where, $F^\prime_{\mu\nu}$ is the field strength tensors of
$U(1)_D$, and the
covariant derivative is given by
\begin{align}
  D_\mu S &= \left(\partial_\mu + i g_D A_\mu^\prime\right) S \,.
\end{align}
Here, $g_D$ is the gauge coupling. And, $A_\mu^\prime$ is the gauge boson
of $U(1)_D$, whose mass is given by $m_D=g_D v_s$ with $v_s$ being obtained from minimize the tree level potential  
\begin{align}
  V_\text{tree}(S) &= - \mu_S^2 S^\dag S
  + \frac{\lambda_S}{2} (S^\dag S)^2\;,
  \label{eq:Vtree}
\end{align}
through
\begin{align}
~~\frac{dV_\text{tree}(S)}{ds}|_{s=v_s}=0\;.
\end{align}
%

For the scalar explanation of the $(g-2)_\mu$ anomaly, we consider relevant new Yukawa couplings for CP-even ($y^{s}$ ) and CP-odd scalars ($y^{p}$) are
      \begin{equation}
      {\cal L} \ \supset \ s
      \left( y_{ij}^s \phi \bar{l_i}l_j + y_{ij}^p \phi \bar{\l_i}\gamma_5\l_j \right)\;,
      \label{eq:s-FNU}
      \end{equation}
Thus, the contribution of a neutral scalar $S$ to the triangle one-loop diagram of $g-2$ is~\cite{Lindner:2016bgg}
\begin{eqnarray}
\Delta a_{\ell}\ = \ \frac{|y_{i j}|^{2}}{8\pi^{2}}
\frac{m_{\ell}^{2}}{m_{S}^{2}}\int_{0}^{1}{\rm d}x\frac{x^{2}(1-x \pm \epsilon_{l_i})}
{(1-x)(1-x\epsilon_{S}^{2})+x\epsilon_{l_j}^{2}\epsilon_{S}^{2}}\,,
\label{eq:g-2_s}
\end{eqnarray}
Here, the plus (minus) sign in the numerator represents the case of CP-even (CP-odd) neutral scalar, and we adopt $\epsilon_{S}\equiv m_{\ell}/m_{S}$ with $m_S$ being the neutral scalar mass. Note that for both gauge and scalar mediators the charged lepton flavors $l_i$ and $l_j$ can be the same or different.
For a complete model, one can consider the coupling of $y_{\ell\ell^{\prime}}$ results from 
integrate out heavy degree freedoms of the SM. For example, the SM Higg doublet interact with the dark $Z_3$ scalar model through the scalar potential as,
\begin{eqnarray}
  V &= &\mu_{H}^{2} (H^\dagger H) + \lambda_{H} (H^\dagger H)^2+ \lambda_{SH} (S^\dagger S) (H^\dagger H) 
  \nonumber\\
  &&+ \mu_{S}^{2} S^{\dagger} S + \lambda_{S} (S^\dagger S)^2 + \frac{\mu_3}{2} (S^{3} + S^{\dagger 3})\;.
\end{eqnarray}
Where,
\begin{equation}
  H =
  \begin{pmatrix}
    G^+
    \\
    \frac{ h + i G^0}{\sqrt{2}}
  \end{pmatrix},
  \qquad
  S = \frac{ s + i \chi}{\sqrt{2}}.
\end{equation} 
The cubic $\mu_{3}$ term breaking the global $U(1)$ $S \to e^{i \alpha} S$ symmetry, and yields a remanent unbroken $Z_{3}$ symmetry.
For the case where the interaction between the visible and dark sector is weak enough, the visible sector will not affect the phase transition in the dark sector at low scales.

\section{GW from dark FOPT}
\label{sec:gw}

As a preparation to investigate the possibility to explain the 12.5-yr NANOGrav result, we here consider the low-scale phase transition.
Four crucial parameters for the estimation of the GW from FOPT are: phase transition temperature $T_\star$, phase transition strength $\alpha$, the inverse duration of the phase transition, and the bubble wall velocity. We take $\alpha=\frac{\Delta\rho}{\rho_R}\;$ with the $\rho_R$ and $\Delta \rho$ being the radiative energy in the plasma and the released latent heat. The inverse time duration of the phase transition is obtained as $
\frac{\beta}{H_n}=T\frac{d (S_3(T)/T)}{d T}|_{T=T_n}\; $.
For magnification of the GW, we take the wall velocity $v_b\approx 1$ with the calculation method follows Ref.~\cite{Caprini:2015zlo}. In addition, for a large $\alpha$, the duration of the phase transition would be very important~\cite{Ellis:2019oqb,Ellis:2020awk,Caprini:2019egz,Ellis:2018mja,Guo:2020grp}. For more details on phase transition and gravitational wave calculations, see Appendix.
 
\section{Numerical results}

\begin{figure}[!htp]
\begin{center}
\includegraphics[width=0.4\textwidth]{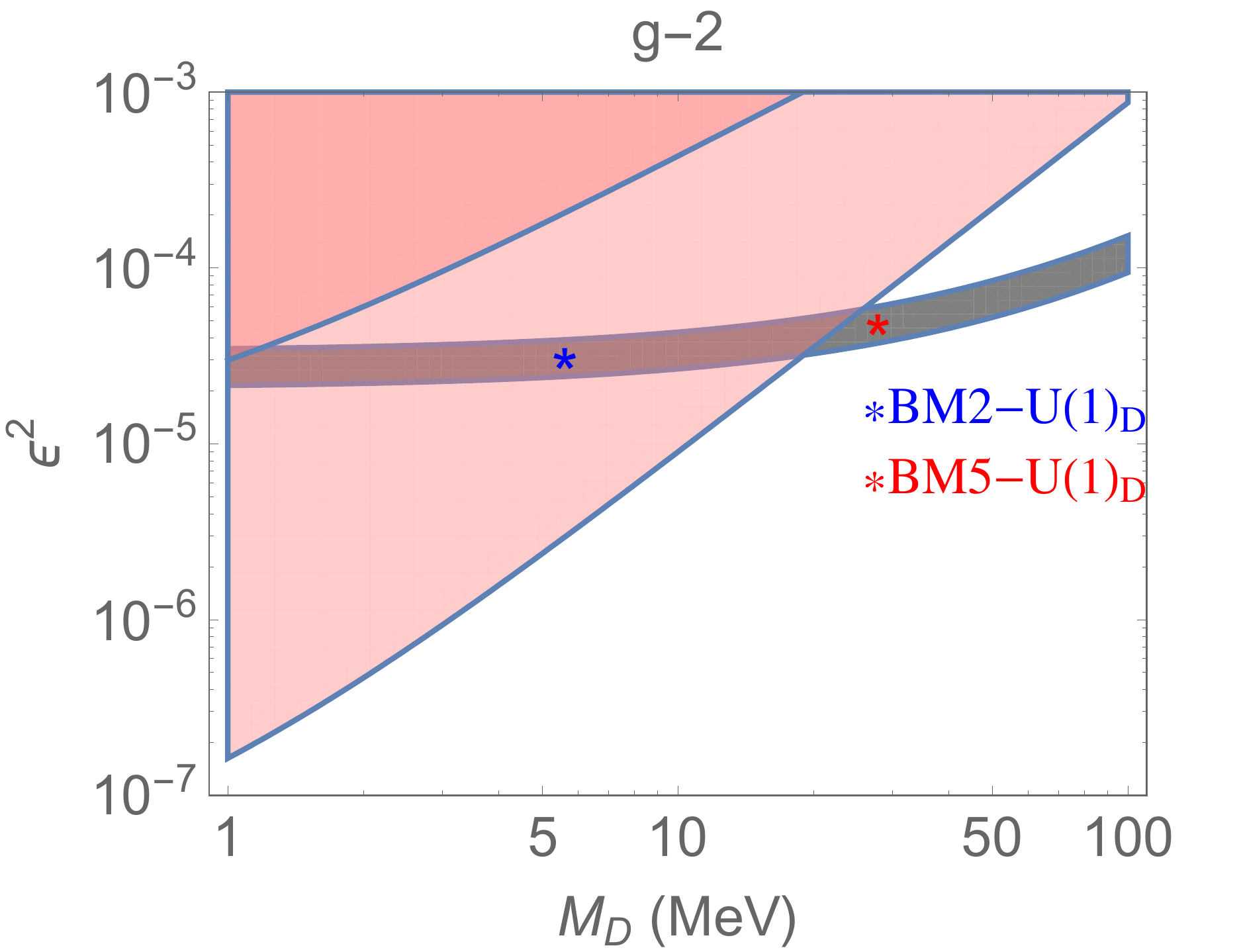}
\includegraphics[width=0.4\textwidth]{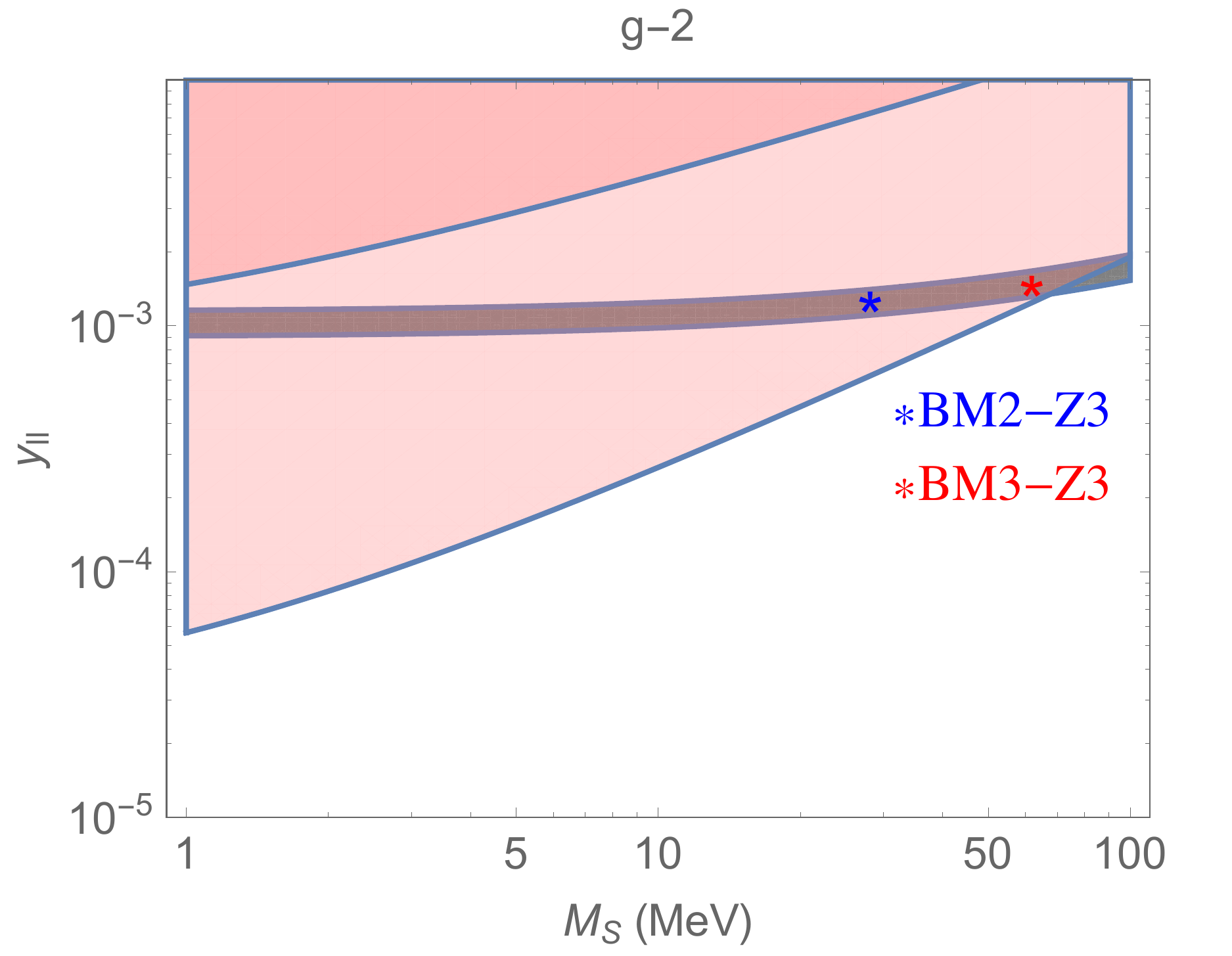}
\caption{The $\Delta a_\mu$ $1\sigma$ parameter regions in the U(1) model and $Z_3$ scalar models.  The darker pink and the lighter pink represent $(g-2)_e$ experiment excluded regions for the case of $r_{V,S}=\sqrt{3/2}$ and $r_{V,S}=\sqrt{3/2}$.
 }\label{ptgwg2}
\end{center}
\end{figure}

We first present in Fig.~\ref{ptgwg2} the possibility to address the $\Delta a_\mu$ anomaly with dark U(1) gauge boson and the $Z_3$ complex scalar model, 
where the $(g-2)_e$ exclude a lot of parameter spaces of light gauge or scalar masses, see pink regions.  Generally, the $(g-2)_e$ bound can be avoided when the dark gauge boson (or $Z_3$ scalar) couple with the $e$ and $\mu$ with a ratio~\cite{Dev:2020drf}: $r_V=\sqrt{g_D^e/g_D^\mu}$, (or $r_S=\sqrt{y_{ee}/y_{\mu\mu}}$). The NA64 would exclude more parameter spaces when invisible decay of dark gauge boson is taken into account~\cite{Banerjee:2016tad,NA64:2019imj}. The GWs from FOPT for different benchmarks for dark U(1) and $Z_3$ scalar are marked in the two plots.

\begin{table}[!htp]
\begin{center}
\begin{tabular}{c c c c c c c c ccc }
\hline
BM~&~$g_{D}$~~ & ~$M_D$(MeV)~&~$T_n$(MeV)~&~$\alpha$~ & ~~$\beta/H_n$~ \\
\hline
$BM_1$  & $0.817$ & $6.866$ & $1.101$ & $0.503$ & $142.249$   \\

$BM_2$  & $0.675$ & $5.616$ & $0.660$ & $3.489$ & $217.012$   \\

$BM_3$  & $0.697$ & $3.501$ & $0.438$ & $2.418$ & $285.977$   \\

$BM_4$  & $0.488$ & $2.484$ & $0.384$ & $3.620$ & $950.743$   \\

$BM_5$  & $0.349$ & $27.948$ & $7.215$ & $0.970$ & $3995.246$   \\

$BM_6$  & $0.559$ & $27.971$ & $17.152$ & $0.013$ & $3393.761$   \\
\hline
\end{tabular}
\caption{The six benchmark points for GW in Fig.~\ref{ptgw}.}
\label{tabp1}
\end{center}
\end{table}

\begin{figure}[!htp]
\begin{center}
\includegraphics[width=0.22\textwidth]{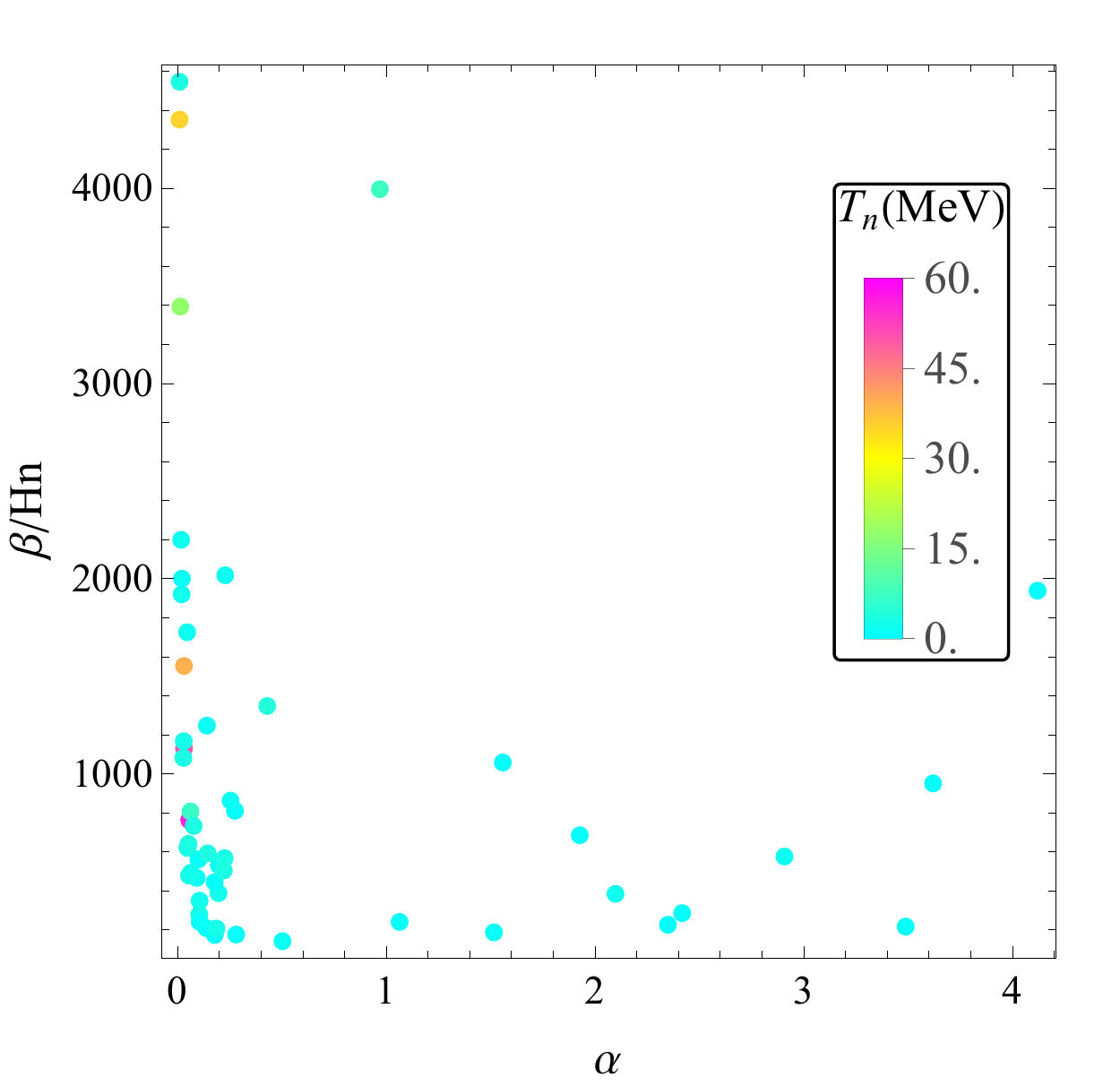}
\includegraphics[width=0.21\textwidth]{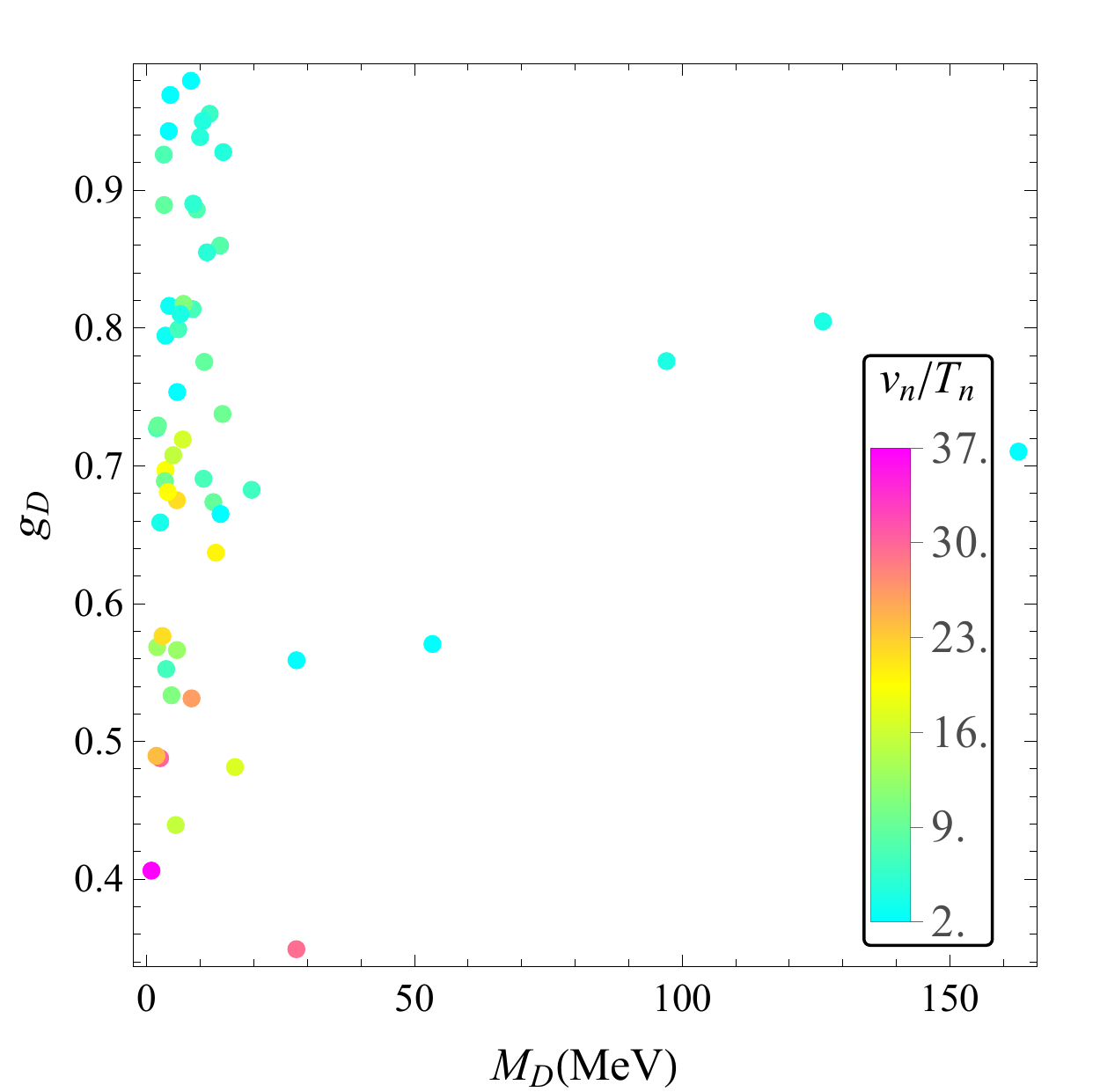}
\includegraphics[width=0.45\textwidth]{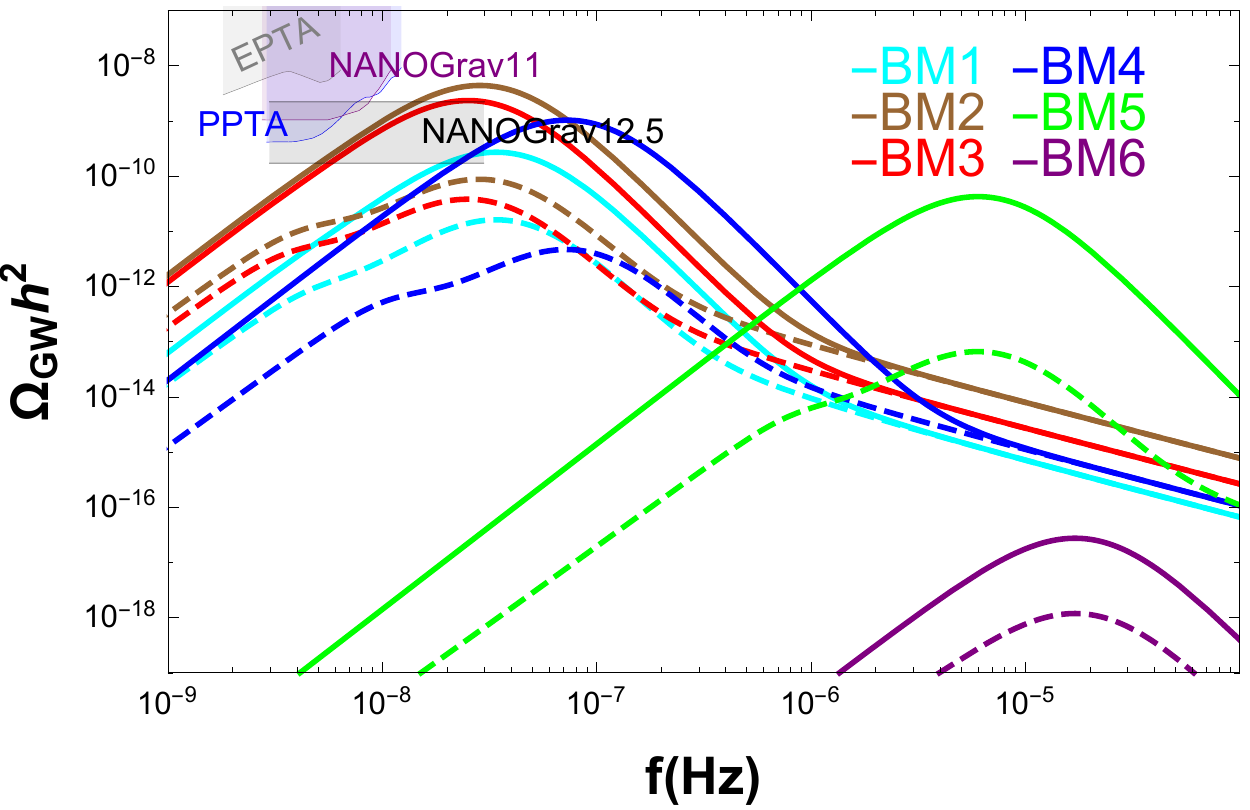}
\caption{FOPT parameter spaces in the dark $U(1)$ model (Top-left); GW parameters of $\alpha_{PT}$, $\beta/H_n$, and $T_n$ in the U(1) model (Top-right); The GW energy density spectrums of low scale FOPT in the dark U(1) model (Bottom).  Dashed (solid) lines represents the case (without) considering the effect of phase transition duration. }\label{ptgw}
\end{center}
\end{figure}

\begin{figure}[!htp]
\begin{center}
\includegraphics[width=0.22\textwidth]{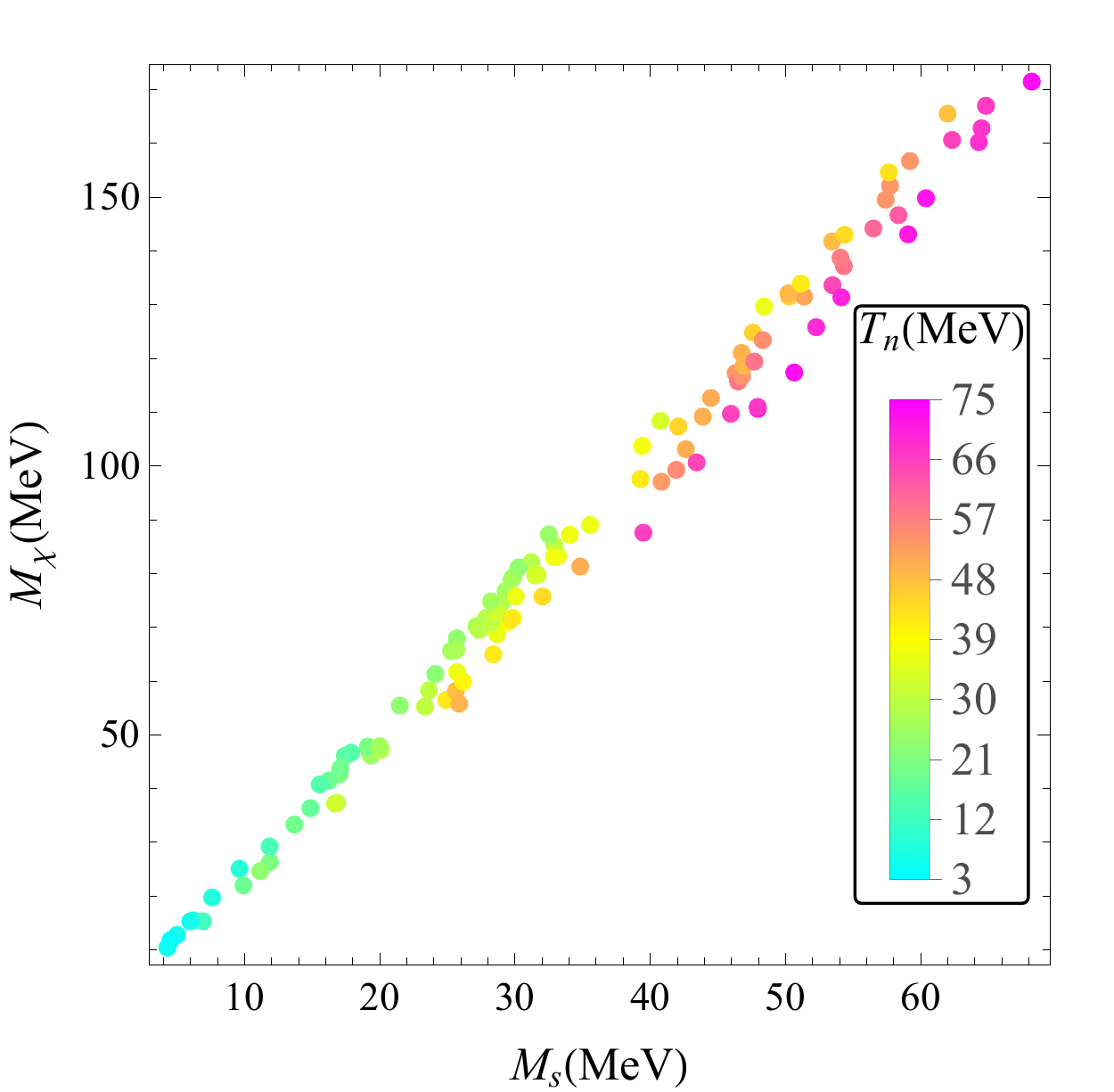}
\includegraphics[width=0.23\textwidth]{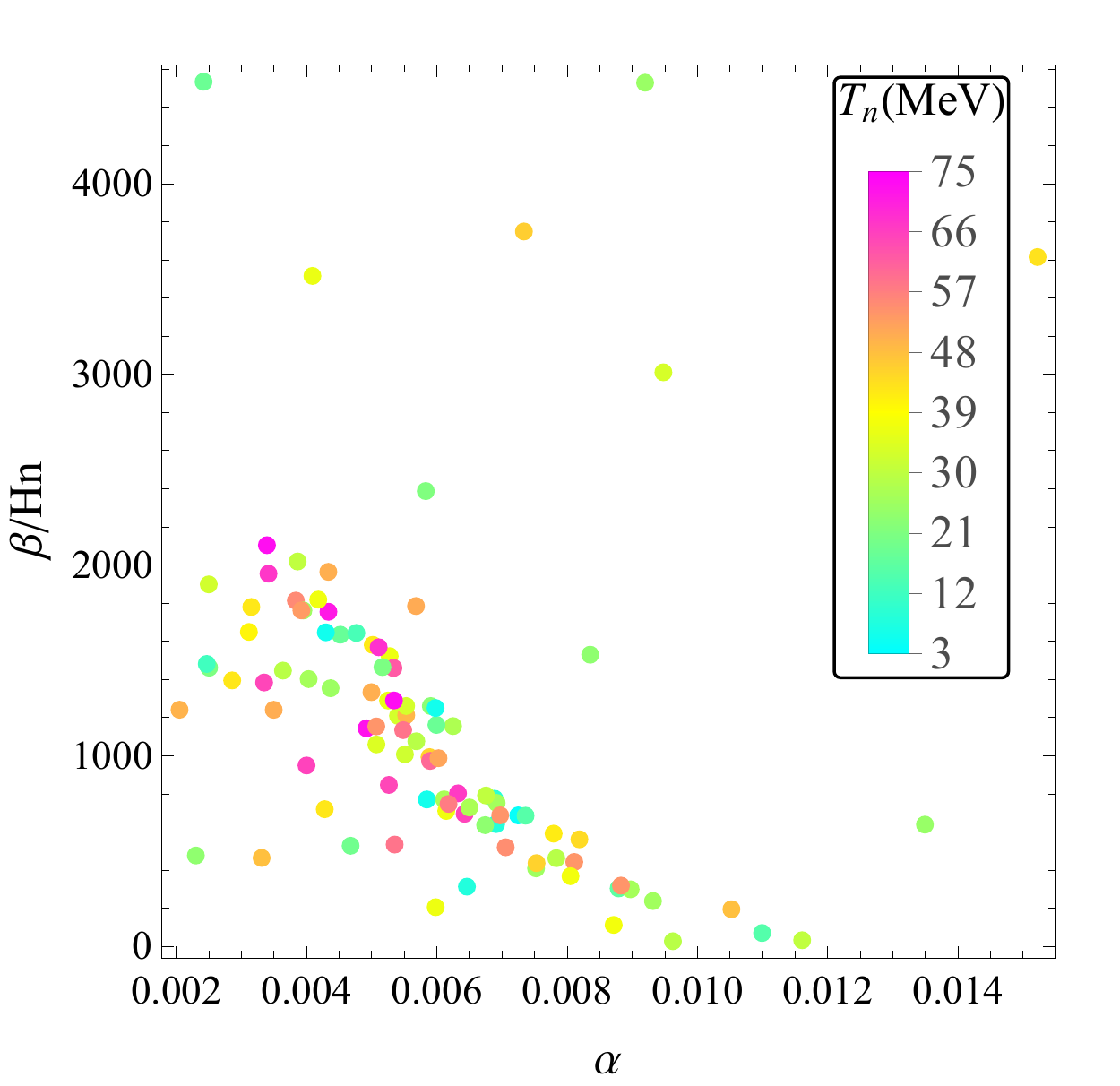}
\includegraphics[width=0.45\textwidth]{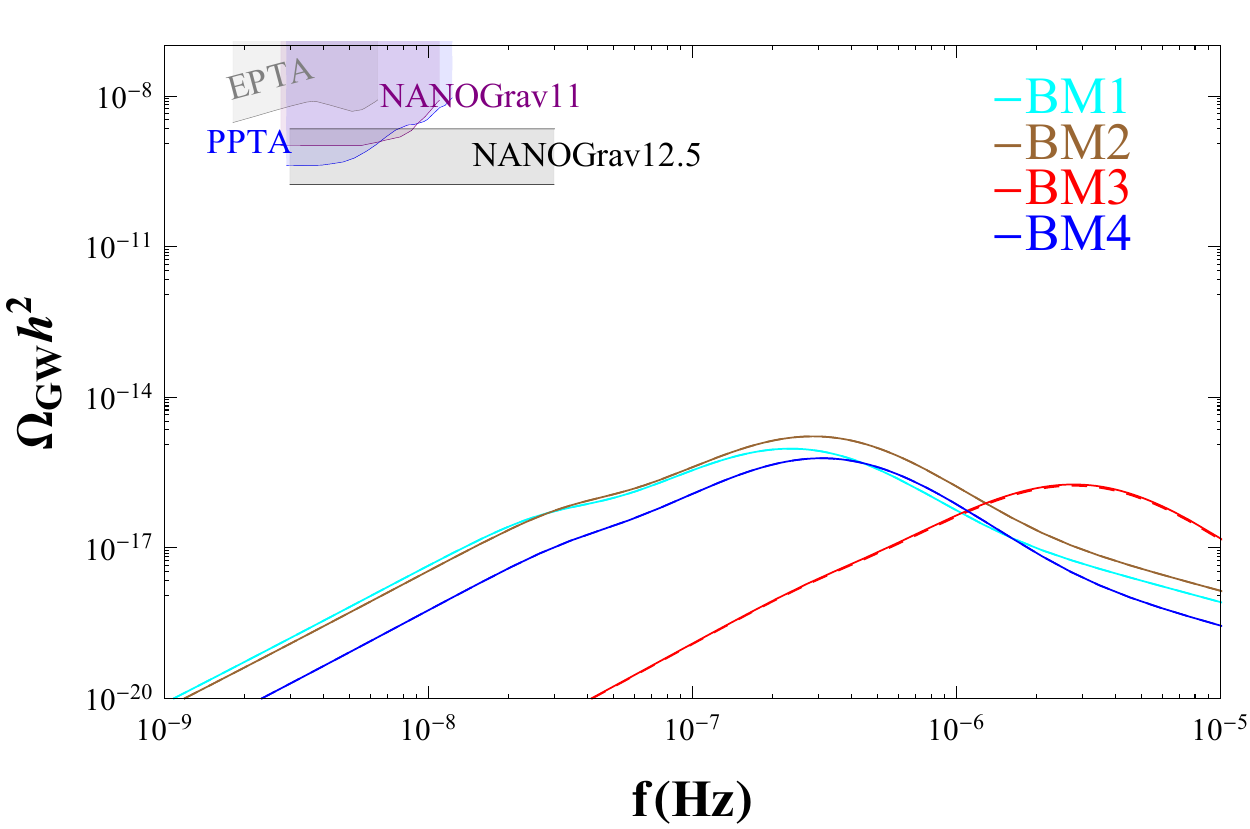}
\caption{FOPT parameter spaces in the dark $Z_3$ model (Top-left); GW parameters of $\alpha_{PT}$, $\beta/H_n$, and $T_n$ in the $Z_3$ model (Top-right);  The GW energy density spectrums of Phase Transition(Bottom). Dashed (solid) lines represents the case (without) considering the effect of phase transition duration. }\label{z3ptgw}
\end{center}
\end{figure}

\begin{table}[!htp]
\begin{center}
\begin{tabular}{c c c c c c c c c c c }
\hline
BM~&~$\lambda_{s}$~~ & ~$M_s$(MeV)~&~$M_\chi$(MeV)~&~$T_n$(MeV)~&~$\alpha$~ & ~~$\beta/H_n$~ \\
\hline
$BM_1$  & $0.369$ & $27.863$ & $71.708$ & $29.146$ & $0.010$ & $26.952$  \\

$BM_2$  & $0.261$ & $28.249$ & $70.833$ & $30.029$ & $0.012$ & $32.176$  \\

$BM_3$  & $0.855$ & $61.994$ & $165.494$ & $47.979$ & $0.011$ & $194.559$  \\

$BM_4$  & $0.445$ & $15.573$ & $40.746$ & $14.853$ & $0.011$ & $69.993$  \\
\hline
\end{tabular}
\caption{The four benchmark points for GW in Fig.~\ref{z3ptgw}.}
\label{tabp2}
\end{center}
\end{table}

Previous studies of the explanation of NANOGrav 12.5-yr data~\cite{Arzoumanian:2020vkk} with stochastic gravitational wave background from FOPT shows that the NANOGrav result bounds the FOPT at temperature $T_n\sim \mathcal{O}(1-100)$ MeV~\cite{Ratzinger:2020koh,Bian:2020bps}. 
This kind of low scale FOPT may occur in the dark sectors weakly interact with the Standard Model~\cite{Breitbach:2018ddu}.
Considering three sources of GWs from FOPT (sound wave, MHD turbulence and collision term), we plot the GW spectra and EPTA, PPTA, NANOGrav11 bounds on the stochastic gravitational waves in Fig.~\ref{ptgw} and Fig.~\ref{z3ptgw}.  
The dark U(1) model allow FOPT with a strong phase transition strength in the parameter space of $M_D\sim MeV$ and $g_D\sim \mathcal{O}(0.1)$. The FOPT strength $\alpha$ in $Z_3$ model is much smaller with the scalar and pseudoscalar $M_{s,\chi}\sim \mathcal{O}(10)$ MeV. 
The current NANOGrav 12.5-yr results could be a confinement for FOPT of the dark U(1) when the dark gauge boson masses are of order of MeV. When the $r_V=\sqrt{3/2}$, these BM points can account for $(g-2)_\mu$ anomaly.  For the case of dark $Z_3$ complex scalar model,  in the parameter spaces that can address the $(g-2)_\mu$ anomaly, the magnitude of GW is too low to be probed by current pulsar timing array experiments.

\section{Conclusions and discussions}
\label{sec:con}

In this paper, we study the possibility to explain the $(g-2)_\mu$ anomaly observed by the FermiLab with dark sectors, including dark $U(1)$ and $Z_3$ complex scalar models. We studied the low mass regions of the parameter spaces that can address the anomaly and found that they are constrained by the $(g-2)_e$ results if the couplings to electron and muon are the same. For the dark $U(1)$ model, we found that the GWB from the FOPT in $U(1)$ maybe serves as an interpretation of the common spectrum process observed by the NANOGrav 12.5-yr dataset. Current pulsar timing array experiments cannot probe the $Z_3$ scalar due to the much lower magnitude of the GWB spectrum.

\section{Acknowledgements}

This work is supported by the National Natural Science Foundation of China under grant No.11605016 and No.11947406, and the Fundamental Research Funds for the Central Universities of China (No. 2019CDXYWL0029).
J.S. is supported by the National Natural Science Foundation of China under Grants No.12025507, No.11690022, No.11947302; and is supported by the Strategic Priority Research Program and Key Re- search Program of Frontier Science of the Chinese Academy of Sciences under Grants No.XDB21010200, No.XDB23010000, and No. ZDBS-LY-7003. 

\appendix
\section{The Phase Transition models}
\label{sec:model}

With the standard methodology, the phase transition can be studied with the thermal one-loop effective potential~\cite{Quiros:1999jp},
\begin{align}
V_{eff}(s,T)=V_{\rm 0} (s)+ V_{CW}(s)+V^{\rm c.t}_{1}(s)+ V_{1}^{T}(s,T)\, .
\end{align}
The Coleman-Weinberg contribution is given by~\cite{Coleman:1973jx}	
\begin{align}
	V_{CW}(s)= \sum_{i} \frac{g_{i}(-1)^{F}}{64\pi^2}  m_{i}^{4}(s)\left(\mathrm{Ln}\left[ \frac{m_{i}^{2}(s)}{\mu^2} \right] - C_i\right)\,,
	\label{eq:oneloop}
\end{align}
Where,  $F=0 \; (1)$ for bosons (fermions), $\mu$ is the $\overline{\text{MS}}$ renormalization scale, $g_{i}$ represent degree of freedoms, $C_i=5/6$ for gauge bosons and $C_i=3/2$ for scalar fields and fermions.
To prevent shifts of the masses and VEVs of the scalars from their tree level values, we impose
\begin{eqnarray}
\partial_{s} (V_{CW}(s) + V^{\rm c.t}_{1}(s)) \bigg|_{s=v_s}&=& 0\,,\nonumber\\
\partial_{s} \partial_{s} (V_{CW}(s) + V^{\rm c.t}_{1}(s)) \bigg|_{s=v_s}&=& 0\,.
\end{eqnarray}

The finite temperature effective potential at one-loop level is given by
\begin{eqnarray}
V_{1}^{T}(s, T)= \frac{T^4}{2\pi^2}\, \sum_i g_i J_{B,F}\left( \frac{ M_i^2(s)+\Pi_i(T)}{T^2}\right),
\end{eqnarray}
 where the Debye masses are calculated as
\begin{align}
  \Pi_{S(G)}(T)  = \left(\frac{\lambda_S}{6} + \frac{g_D^2}{4}\right) T^2 \,, ~~~
  \Pi_{A'}(T) = \frac{g_D^2}{3} T^2 \,.
\end{align}
The functions $J_{B,F}(y)$ are
\begin{eqnarray}
 J_{B,F}(y) = \pm \int_0^\infty\, dx\, x^2\, \ln\left[1\mp {\rm exp}\left(-\sqrt{x^2+y}\right)\right]\;.
\end{eqnarray}
Where, the upper (lower) sign corresponds to bosonic (fermionic) contributions. The above integral $J_{B,F}$ can be expressed as a sum of the second kind modified Bessel functions $K_{2} (x)$~\cite{Bernon:2017jgv},
\begin{eqnarray}
J_{B,F}(y) = \lim_{N \to +\infty} \mp \sum_{l=1}^{N} {(\pm1)^{l}  y \over l^{2}} K_{2} (\sqrt{y} l)\;.
\end{eqnarray}

For the $U(1)_D$ model, we have the tree-level potential for the classical scalar field being 
\begin{equation}
V_{0} (s) =- \frac{\mu_s^2}{2}s^2 + \frac{\lambda_s}{8} s^4.
\end{equation}
For the calculation of the Coleman-Weinberg potential and the $V_{1}^{T}(s, T)$, we need the field-dependent masses of the scalar, Goldstone, and of the gauge boson, that are given by
\begin{align}
  &m^2_{S}(s)   = -\mu_S^2+\frac{3}{2}\lambda_S s^2 \,,~~
  m^2_{G_D}(s) = -\mu_S^2+\frac{1}{2}\lambda_S s^2 \,,\nonumber\\
  &m^2_{A'}(s)       = g_D^2 s^2 \,.
\end{align}
The counter terms to the potential in Eq.~\eqref{eq:oneloop} is obtained as
\begin{equation}
V^{\rm c.t}_{1} (s) = -\frac{\delta\mu_2^2}{2} s^{2} + \frac{\delta\lambda_2}{8}s^{4}.
\label{eq:CT}
\end{equation}

For the $Z_3$ scalar model, the tree-level potential in terms of the classical field is
 \begin{align}
 V_\text{0}(s)=\frac{\mu_s^{2}}{2}s^{2}+\frac{\mu_3}{2 \sqrt{2}}s^{3}+\frac{\lambda_{S}}{4}s^{4}\;.
 \end{align}
The counter-terms are calculated as
\begin{align}
  V_\text{ct}(s) &=  \frac{\delta\mu_s^2}{2} s^2
+ \frac{\delta\mu_3}{2\sqrt{2}} s^3
+ \frac{\delta\lambda_s}{4} s^4\;,
\end{align}
and can be fixed through following conditions,
\begin{align}
& \frac{d (V_\text{CW}(s)+ V_\text{ct}(s))}{d s}\bigg|_{s=v_s} = 0 \,,
 \frac{d^2 (V_\text{CW}(s)+ V_\text{ct}(s)) }{d s^2}\bigg|_{s=v_s}= 0 \,, \nonumber\\
& V_\text{CW}(0) - V_\text{CW}(v_s)- V_\text{ct}(v_s)= 0 \,.
\end{align}
Here, the thermal masses of the scalar degree of freedoms are
\begin{align}\label{eqms}
  m_s^2(s,T) &= m_s^2+\frac{\lambda_S}{3} T^2 \,, \\
  m_\chi^2(s,T) &= m_\chi^2+\frac{\lambda_S}{3} T^2 \,.
\end{align}

With the thermal effective potential at hand, one can get
the solution of the bounce configuration of the nucleated bubbles through extremizing,
\begin{eqnarray}
S_3(T)=\int 4\pi r^2d r\bigg[\frac{1}{2}\big(\frac{d \phi_b}{dr}\big)^2+V_{eff}(\phi_b,T)\bigg]\;,
\end{eqnarray}
after solving the equation of motion for the classical field $\phi_b$,
by considering the following boundary conditions
\begin{eqnarray}
\lim_{r\rightarrow \infty}\phi_b =0\;, \quad \quad \frac{d\phi_b}{d r}|_{r=0}=0\;.
\end{eqnarray}
The $\phi_b$ is the $s$ field considered in this work.
The onset of the phase transition is characterize by:
\begin{eqnarray}\label{eq:bn}
\Gamma\approx A(T)e^{-S_3/T}\sim 1\;,
\end{eqnarray}
when the number of bubbles for bubble nucleation per horizon volume and per horizon time is of order unity, one can obtain the nucleation temperature ($T_n$)~\cite{Affleck:1980ac,Linde:1981zj,Linde:1980tt}.

After the analysis of the FOPT, one can get the phase transition temperature, the phase transition strength, and the phase transition duration that are crucial for the calculation of GWs from the FOPT. 
The gravitational waves from the FOPT come from bubble collision, sound waves, and the MHD turbulence~\cite{Caprini:2015zlo}.  In which, generally, the sound waves constitutes the leading source of GWs,
whose energy density is
\begin{eqnarray}
\Omega h^2_{\rm sw}(f)&=&2.65 \times 10^{-6}(H_*\tau_{sw})\left(\frac{\beta}{H}\right)^{-1} v_b
\left(\frac{\kappa_\nu \alpha }{1+\alpha }\right)^2\nonumber\\
&\times&\left(\frac{g_*}{100}\right)^{-\frac{1}{3}}
\left(\frac{f}{f_{\rm sw}}\right)^3 \left(\frac{7}{4+3 \left(f/f_{\rm sw}\right)^2}\right)^{7/2},
\end{eqnarray}
where, the duration of the phase transition is characterized by the factor of $\tau_{sw}=min\left[\frac{1}{H_*},\frac{R_*}{\bar{U}_f}\right]$, $H_*R_*=v_b(8\pi)^{1/3}(\beta/H)^{-1}$~\cite{Ellis:2020awk}. The factor of  
\begin{equation}
\bar{U}_f^2\approx\frac{3}{4}\frac{\kappa_\nu\alpha}{1+\alpha}\;
\end{equation}
is the root-mean-square (RMS) fluid velocity~\cite{Hindmarsh:2017gnf, Caprini:2019egz, Ellis:2019oqb}.
The term $H_*\tau_{\rm sw}$ is to account for the suppression effect  of the phase transition duration for the GW amplitude of the sound wave source. The $\kappa_\nu$ factor is the fraction of the latent heat transferred into the kinetic energy of plasma, which is given by the hydrodynamic analysis~\cite{Espinosa:2010hh}.
The peak frequency of the sound wave locates at~\cite{Hindmarsh:2017gnf}
\begin{equation}
f_{\rm sw}=1.9 \times 10^{-5} \frac{\beta}{H} \frac{1}{v_b} \frac{T_*}{100}\left({\frac{g_*}{100}}\right)^{\frac{1}{6}} {\rm Hz }\;.
\end{equation}
In this study, we take $T_*=T_n$.

The gravitational waves from the bubble collision can be obtained with the envelop approximation~\cite{Kosowsky:1991ua,Kosowsky:1992rz,Kosowsky:1992vn} is~\cite{Huber:2008hg},
\begin{eqnarray}
  \Omega_{\text{col}} h^2 =& 1.67\times 10^{-5} \left(\frac{H_{\ast}}{\beta}\right)^2
  \left(\frac{\kappa \alpha}{1+\alpha}\right)^2 \left( \frac{100}{g_{\ast}} \right)^{1/3}\nonumber\\
  &\left( \frac{0.11 v_b^3}{0.42 + v_b^2} \right)
  \frac{3.8(f/f_{\text{env}})^{2.8}}{1+2.8(f/f_{\text{env}})^{3.8}} \; .
\end{eqnarray}
Here, $v_b$ is the bubble wall velocity,  the efficient factor $\kappa$ which characterizes the fraction of latent heat
deposited in a thin shell. We take $\alpha$ as a function of the phase strength parameter $\alpha$~\cite{Kamionkowski:1993fg}:
\begin{eqnarray}
\kappa \simeq  \frac{0.715\alpha + \frac{4}{27} \sqrt{3\alpha/2}}{1+0.715\alpha} \;,
\end{eqnarray}
the peak frequency $f_{\text{env}}$ is given by,
\begin{eqnarray}
  f_{\text{env}} = 16.5\times 10^{-6} \left(\frac{f_{\ast}}{H_{\ast}}\right)  \left(\frac{T_{\ast}}{100\text{GeV}}\right)
  \left(\frac{g_{\ast}}{100}\right)^{1/6}
  \text{Hz}\; . \nonumber
\end{eqnarray}

The MHD turbulence in the plasma might be another important source of the gravitational wave from FOPT though it is still suffer from large uncertainty~\cite{Caprini:2019egz}, whose the peak frequency locates at \cite{Caprini:2009yp}
\begin{equation}
f_{\rm turb}=2.7  \times 10^{-5}
\frac{\beta}{H} \frac{1}{v_b} \frac{T_*}{100}\left({\frac{g_*}{100}}\right)^{\frac{1}{6}} {\rm Hz }\;,
\end{equation}
 the energy density spectrum is given by
\begin{eqnarray}
\Omega h^2_{\rm turb}(f)&=3.35 \times 10^{-4}\left(\frac{\beta}{H}\right)^{-1}
\left(\frac{\epsilon \kappa \alpha }{1+\alpha }\right)^{\frac{3}{2}}
\left(\frac{g_*}{100}\right)^{-\frac{1}{3}}
\nonumber \\ &\times v_b
\frac{\left(f/f_{\rm turb}\right)^3\left(1+f/f_{\rm turb}\right)^{-\frac{11}{3}}}{\left[1+8\pi f a_0/(a_* H_*)\right]}\;.
\end{eqnarray}
Where, the precent Hubble parameter is
\begin{equation}
	h_{\ast} = \bigl( 1.65 \times 10^{-5} Hz \bigr) \left( \frac{T_{*}}{100 \rm{GeV}} \right) \left( \frac{g_{\ast}}{100} \right)^{1/6}\;.
\end{equation}
We take efficiency factor $\epsilon \approx 0.05$ for the GW calculations.

\bibliographystyle{unsrt}

\end{document}